\documentstyle[12pt,subeq]{article}

\begin{document}

\title{Boundary behavior of quantum Green's Functions}

\author{
L. {\v S}amaj$^{1,2}$, J. K. Percus$^{2,3}$ and P. Kalinay$^1$
}

\maketitle

\begin{abstract}
We consider the time-independent Green function for the Schr\"odinger 
operator with a one-particle potential, defined in a
$d$-dimensional domain. 
Recently, in one dimension (1D), the Green's function problem
was solved explicitly in inverse form, with diagonal
elements of Green's function as prescribed variables.
In this paper, the 1D inverse solution is used to derive 
leading behavior of Green's function close to the domain boundary.
The emphasis is put onto "universal" expansion terms which are 
dominated by the boundary and do not depend 
on the particular shape of the applied potential.
The inverse formalism is extended to higher dimensions, especially to 3D, 
and subsequently the boundary form of Green's function is predicted for 
an arbitrarily shaped domain boundary.
\end{abstract}

\vfill

\noindent $^1$ Institute of Physics, Slovak Academy of Sciences,
D\'ubravsk\'a cesta 9, 842 28 Bratislava, Slovakia;
E-mail: fyzimaes@savba.sk 

\noindent $^2$ Courant Institute of Mathematical Sciences,
New York University, 251 Mercer Street, New York, NY 10012

\noindent $^3$ Physics Department, New York University,
New York, NY 10003

\newpage

\renewcommand{\theequation}{1.\arabic{equation}}
\setcounter{equation}{0}

\section{Introduction}
For a connected finite domain, the eigenvalue spectrum
of an operator with a boundary condition depends strongly
on the shape of the smooth boundary.
A rigorous analysis of the eigenvalue density was first
done by Weyl$^1$ for the 3D Laplacian in the asymptotic limit of an 
infinite volume (at short wavelengths of eigenstates).
The next surface and curvature contributions to the eigenvalue density
were derived for the Dirichlet and Neumann boundary conditions
first in 3D,$^2$ and then, by using the path integral
method, in 2D.$^3$
In this connection, Kac posed a question whether there is a one-to-one
correspondence between the boundary shape and the corresponding
eigenvalue spectrum of the Laplacian.
The answer is negative:$^4$ there exist nonisometric pairs
of 2D flat shapes which are isospectral.$^{5,6}$ 
An important progress in calculations was made within
a time-independent Green function method in Ref. 7.
Here, a multiple reflection expansion was used to get Green's function
for an arbitrary domain, in the limit of small wavelengths.
This technique, applied to physical problems like electromagnetic
field in a cavity,$^8$ evoqued numerous research 
activities$^9$.
Its extension to the Schr\"odinger operator$^{10}$ resulted in 
semiclassical expansions for quantum mechanical system 
valid for finite domains$^{11,12}$.
The influence of spatial confinement on the energy spectra
of simple quantum systems, like harmonic oscillator
or hydrogen atom, was investigated in many other works
(see e.g. Refs. 13-17).

This paper deals with local effects of a domain
boundary on Green's function of a Schr\"odinger operator. 
The topic is studied within an "inverse" formulation for 
Green's functions.
The motivation for this formulation comes from the seemingly
unrelated density-functional theory.
The latter is based on the famous Hohenberg-Kohn uniqueness
theorem,$^{18}$ having a practical realization in the
Kohn-Sham separation ansatz.$^{19}$
In the density-functional theory, the usual "direct" formulation,
i.e. the ground-state expectation value of an observable in terms
of some prescribed external fields, is replaced by the "inverse"
one, with the particle density as the controlling variable.
In a series of works,$^{20-22}$ we have developed an infinite
gradient series expansion of the kinetic-energy density functional
for 1D noninteracting Fermion systems in the ground state 
by applying the Green's function method.
By definition, the expansion in potential gradients and correspondingly
in density gradients does not allow us to consider discontinuities,
say at hard particles or extended regions.
The "inverse" formulation of Green's functions arised as a natural 
mean for their explicit representation in any situation. 

To introduce the inverse format for Green's functions,
let us first specify notation.
We consider a $d$-dimensional domain $\Omega$ of points
${\bf r} = (x_1,x_2,\ldots,x_d)$, infinite 
$({\bf r} \in R^d)$ or finite, bounded by hard walls
at the domain boundary $\partial \Omega$.
Selecting units so that $\hbar^2/(2m)=1$
with $m$ being the particle mass, the general one-particle
Hamiltonian formulated within the domain $\Omega$
is given by
\begin{equation} \label{1.1}
{\bf \hat H} = - \Delta_{{\bf r}} + u({\bf r}) .
\end{equation}
Here, $u({\bf r})$ is an arbitrary external potential,
continuous inside $\Omega$,  
which does not include the infinite potential due to 
the presence of domain walls at $\partial\Omega$.
The time-independent Schr\"odinger equation reads
\begin{equation} \label{1.2}
{\bf \hat H} \psi_k({\bf r}) = \lambda_k \psi_k({\bf r})
\end{equation} 
with the Dirichlet b.c. $\psi_k({\bf r}) = 0$
at the domain boundary ${\bf r}\in \partial \Omega$.
$\psi_k({\bf r})$ and $\lambda_k$ are respectively
the eigenstates and the eigenvalues of Hamiltonian 
${\bf \hat H}$, and index $k$ can be either
discrete (localized eigenstates) or continuous 
(extended eigenstates).
The time-independent Green's function is defined by$^{23}$
${\bf G}_z = (z - {\bf \hat H})^{-1}$,
where $z$ is a complex variable with components
$\lambda = {\rm Re}(z)$ and $s = {\rm Im}(z)$.
${\bf G}_z$ is an analytic function of $z$ except of 
those points on the real $z$-axis which correspond
to the eigenvalues of $\hat{\bf H}$: it exhibits simple poles
at discrete eigenvalues of $\hat{\bf H}$ and a branch cut
along fragments of the real $z$-axis which correspond
to the continuous spectrum of $\hat{\bf H}$.
In the latter case, the discontinuity of the transverse limits 
${\bf G}^{\pm}_{\lambda} = \lim_{s\to 0^+} {\bf G}_{\lambda\pm 
{\rm i}s}$ yields the density of states at $\lambda$.
In the vector-space representation, 
\begin{equation} \label{1.3}
G_z({\bf r},{\bf r}')  =  
\sum_k \frac{\psi_k^*({\bf r}) \psi_k({\bf r}')}{z-\lambda_k} ,
\end{equation}
Green's function satisfies the two-point differential equation
\begin{equation} \label{1.4}
\left[ \Delta_{\bf r} + z - u({\bf r}) \right] G_z({\bf r},{\bf r}')
= \delta({\bf r}-{\bf r}')
\end{equation}
with the b.c. at the domain boundary,
$G_z({\bf r},{\bf r}') = 0$ at ${\bf r}\in \partial \Omega$.
We notice that in one dimension (1D),
diagonal elements of Green's function, denoted by
\begin{equation} \label{1.5}
n_z(x) =  \sum_k \frac{\psi_k(x) \psi_k^*(x)}{z-\lambda_k} ,
\end{equation}
are well defined finite quantities.
This is no longer true in dimensions $\ge$ 2 where
the equal-argument Green's function diverges.
One can avoid this divergence by redefining
new one-point quantities, which choice is crucial from the
point of view of this work.
In 3D, which is of practical interest, the finite quantity
\begin{equation} \label{1.6}
\nu_z({\bf r}) = \lim_{{\bf r}' \to {\bf r}} 4 \pi
\frac{\partial}{\partial z} G_z({\bf r},{\bf r}') ,
\quad d=3 ,
\end{equation}
will be especially useful.
The function $\nu_z$ remains informative since the 
differentiation of Green's function with respect to $z$ keeps
the poles of the latter.
One can reconstruct the original Green's function
by using the relation
\begin{equation} \label{1.7}
\int_0^z {\rm d}s~ \nu_s({\bf r}) = 4\pi
\langle {\bf r} \vert \frac{1}{z-{\bf \hat H}} + 
\frac{1}{{\bf \hat H}} \vert {\bf r} \rangle ,
\quad d=3 .
\end{equation}
The $z$-independent term in (\ref{1.7}), which subtracts
the diverging part of the diagonal Green's function element,
has no effect on the density-functional formalism developed
in Refs. 20-22.

The above format is the direct one: for a given constraining
domain $\Omega$ and a given external field $u({\bf r})$, find
the corresponding Green's function.
For continuum 1D space$^{20,21}$ and for simply connected lattice 
structures,$^{22}$ the direct problem was replaced by the inverse one 
with diagonal elements of Green's function as controlling variables: 
(i) the external field in terms of the diagonal elements of
Green's function (the inverse profile relation);
(ii) off-diagonal Green's function elements in terms of
the diagonal ones.
The 1D inverse profile relation was found explicitly in the form
\begin{equation} \label{1.8}
z - u(x) = - \frac{1}{4}\frac{1}{n_z^2(x)} + \frac{1}{4}
\left[ \frac{n'_z(x)}{n_z(x)} \right]^2
- \frac{1}{2} \frac{n''_z(x)}{n_z(x)} ,
\end{equation}
with the standard b.c. $n_z(x)=0$ at $x\in \partial \Omega$.
A prime means the derivative with respect to the argument.
Eq. (\ref{1.8}) with b.c. determines $n_z(x)$ uniquely up to the sign.
When compared to the basic equation (\ref{1.4}) of the direct
format, it has the virtue of being a one-point differential equation 
without the appearance of a $\delta$-function.
As concerns the off-diagonal elements of Green's function,
they can be expressed in terms of the diagonal ones as follows
\begin{equation} \label{1.9}
G_z(x,x') = \left[ n_z(x) \right]^{1/2} \left[ n_z(x') \right]^{1/2}
\exp \left[ {\rm sign} (x-x') \int_{x'}^x 
\frac{{\rm d}s}{2 n_z(s)} \right] .
\end{equation}

As was already mentioned, in this paper we use the inverse formalism 
to study behavior of the Green's function close to the confining 
domain boundary.
In the leading and some higher orders of the distance from the boundary,
the Green's function is shown to exhibit "universal" terms
which are dominated by the boundary and do not depend on
the applied potential. 
These boundary terms are derived first in 1D, and subsequently 
predicted in 3D for an arbitrarily shaped domain boundary.
The results are exact at all wavelengths.

The paper is organized as follows.
In Sec. 2, we derive in the inverse format the boundary form 
of 1D Green's function.
Before extending the inverse formalism to higher dimensions, 
in particular to 3D, we first treat, in the direct format with
the zero potential, two special cases which can be treated exactly 
as one-dimensional.
Systems that are stratified to only one Cartesian coordinate (Sec. 3) 
justify the choice of the one-point quantity $\nu_z$ (\ref{1.6}).
Spherically symmetric systems (Sec. 4) reveal unexpected logarithmic
terms in the boundary expansion of Green's function.
In the latter case, as a by-product of the formalism, we derive
general behavior of Green functions close to the center of
radial symmetry.  
A technique for systematic construction of the inverse
profile relation in an arbitrary dimension and for any potential
is developed in Sec. 5.
As a consequence, the (universal) boundary terms in 3D Green's 
function are suggested for an arbitrarily shaped domain boundary.

\renewcommand{\theequation}{2.\arabic{equation}}
\setcounter{equation}{0}

\section{1D Green's function}
We start with 1D Green's function $G_z(x,x')$ defined in a domain
$\Omega=(0,X)$, and look for its behavior close to a wall,
say as $x,x' \to 0$.
The potential $u(x)$ is assumed to be regular at $x=0$,
and its Taylor expansion for $x>0$ is written as
\begin{equation} \label{2.1}
u(x) = \sum_{n=0}^{\infty} u_n  x^n ,
\quad \quad u_n = \frac{1}{n!} 
\frac{{\rm d}^n u(x)}{{\rm d}x^n}\big\vert_{x=0} . 
\end{equation}
The first coefficient $u_0$ only shifts $z\to z-u_0$, 
so without any loss of generality it is set equal to 0.

Let us suppose that the diagonal element of Green's function
$n_z(x)$, satisfying the b.c. $n_z(0)=0$, also has the Taylor 
series expansion around $x=0$,
\begin{equation} \label{2.2}
n_z(x) = \sum_{n=1}^{\infty} c_n x^n ,
\quad \quad c_n = \frac{1}{n!} 
\frac{{\rm d}^n n_z(x)}{{\rm d}x^n}\big\vert_{x=0} ,
\end{equation}
and insert this expansion into the exact 
inverse profile relation (\ref{1.8}).
The requirement of the vanishing of the prefactors attached to
$x^{-2}, x^{-1}, x^0,$ etc., implies the following sequence 
of equations for the coefficients $\{ c_n \}$:
\begin{subequations}
\begin{eqnarray}
\left( c_1 \right)^2 - 1 & = & 0 , \label{2.3a} \\
c_2 \left[ \left( c_1 \right)^2 - 1 
\right] & = & 0 , \label{2.3b} \\
2 z c_1 + 3 c_3 & = & 0 , \label{2.3c}
\end{eqnarray}
\end{subequations}
etc.
The first Eq. (\ref{2.3a}) tells us that $[n_z'(0)]^2 = 1$.
Actually, we observe in 1D examples with boundaries, like free
particle or harmonic oscillator in bounded well, that
\begin{equation} \label{2.4}
n_z'(0) = - 1 .
\end{equation}
This means that $n_z$ goes to zero at the boundary from below.
The universal slope $-1$ depends neither on details of 
the applied potential nor on the b.c. at the opposite
boundary $X$.
Relation (\ref{2.4}) also tells us that the functional series,
which determines $n_z(x)$ via formula (\ref{1.5}), is not
uniformly convergent.
In the opposite case, we could differentiate this functional
series term by term with respect to $x$, which leads to
the contradiction $n'_z(0)=0$.
The second Eq. (\ref{2.3b}) is fulfilled identically,
leaving the coefficient $c_2$ unspecified: the value
of $c_2(z)$, which depends on $z$ and $\{ u_n \}$, is fixed by 
the requirement of the vanishing of $n_z(x)$ at the opposite 
$X$-boundary.
The third Eq. (\ref{2.3c}) determines the coefficient $c_3$, which,
being the function of only $z$ (trivially shifted by $u_0$), 
is universal in the same sense as $c_1$.
The next coefficients $c_4,  c_5, \cdots$ are analytic
functions of both $\{ u_n \}$ and $c_2$, which confirms 
the adequacy of the analyticity assumption (\ref{2.2}).
We conclude that
\begin{equation} \label{2.5}
n_z(x) \sim - x + c_2(z) x^2 + \frac{2 z}{3} x^3
\quad \quad {\rm as}\ x\to 0^+ .
\end{equation}

When the two points $x$ and $x'$ are close to the boundary
at $0$, the leading terms of $G_z(x,x')$ can be evaluated
by inserting (\ref{2.5}) into (\ref{1.9}), with the result
\begin{equation} \label{2.6}
G_z(x,x') \sim - x_< + c_2(z) x x' + \frac{z}{6}
x_< \left( x_<^2 + 3 x_>^2 \right)
\quad \quad {\rm as\ both\ } x,x' \to 0^+ .
\end{equation}
Here, we have introduced the standard notation
$x_< \equiv \min\{ x,x' \}$ and $x_> \equiv \max\{ x,x' \}$.

\renewcommand{\theequation}{3.\arabic{equation}}
\setcounter{equation}{0}

\section{d-dimensional Green function with a 1D confinement
to potential}
There exists a large family of $d$-dimensional models whose
Green's function can be explicitly expressed
in terms of a related 1D Green's function. 
The family is defined by potentials that are stratified in
one dimension, i.e. depend on only one Cartesian coordinate 
of ${\bf r} = (x_1,\cdots,x_d)$, say $x_1=x\in \Omega_x$, 
$u({\bf r})=u(x)$.
For simplicity, the subspace of vectors 
${\bf r}_{\perp} = (x_2,\cdots,x_d)$ 
perpendicular to the $x$-axis will be infinite, 
$\Omega = \Omega_x \otimes R^{d-1}$.
Using the separation-of-variables treatment with plane waves 
${\bf k}_{\perp} = (k_2,\ldots,k_d)$ in the
perpendicular subspace, the $d$-dimensional Green's function is 
expressible as
\begin{eqnarray} \label{3.1}
G_z({\bf r},{\bf r}') = \int_{-\infty}^{\infty} \frac{{\rm d}k_2}{2\pi} 
\cdots \int_{-\infty}^{\infty} \frac{{\rm d}k_d}{2\pi}
G_{z-k_2^2-\cdots -k_d^2}^{1{\rm D}}(x,x')
{\rm e}^{{\rm i}{\bf k}_{\perp}
\cdot ({\bf r}_{\perp}-{\bf r}_{\perp}')} ,
\end{eqnarray}
where ${\bf G}^{1{\rm D}}$ is 1D Green's function associated
with the potential $u(x)$ and the domain $\Omega_x$. 
In $d=3$ dimensions, passing to polar coordinates, 
formula (\ref{3.1}) takes form
\begin{equation} \label{3.2}
G_z^{3{\rm D}}({\bf r},{\bf r}') = \frac{1}{2\pi}
\int_0^{\infty} {\rm d}k~ k G_{z-k^2}^{1{\rm D}}(x,x')
J_0(k \rho_{\perp}) .
\end{equation}
Here, $\rho_{\perp} = \vert {\bf r}_{\perp} - {\bf r}'_{\perp} \vert$
is the perpendicular distance of points ${\bf r}, {\bf r}'$ and 
$J_0$ denotes the Bessel function of the first kind.$^{24}$

Let the 3D domain $\Omega$ be the half-space
$\langle 0,\infty) \otimes R^2$.
The corresponding 1D Green's function $G^{1{\rm D}}_{z-k^2}(x,x')$
in (\ref{3.2}), formulated within $\Omega_x = \langle 0,\infty)$,
can be expanded in $x$ and $x'$ near the rectilinear hard wall
at $0$ using the previously derived formula (\ref{2.6}).
It might be tempting to insert this expansion directly into (\ref{3.2}),
but already the leading term $\sim - x_<$ implies a diverging
integral over $k$.
This indicates that a renormalization procedure is needed
for 1D Green's function.

We first assume that $\rho_{\perp}\ne 0$, i.e. the two points
${\bf r}$ and ${\bf r}'$ do not lie on the same line perpendicular
to the surface of the wall.
$G^{1{\rm D}}$ can be split into two parts
\begin{equation} \label{3.3}
G_z^{1{\rm D}}(x,x') = \frac{1}{2{\rm i}\sqrt{z}} \left[
{\rm e}^{{\rm i}\sqrt{z}\vert x-x'\vert} -
{\rm e}^{{\rm i}\sqrt{z}(x+x')} \right] + \delta G_z^{1{\rm D}}(x,x') ,
\quad \quad x,x'\ge 0 ,
\end{equation}
where one assumes that ${\rm Im}(\sqrt{z})>0$ in order to ensure
the regularity at $\vert x-x' \vert \to \infty$.
The first part, which is nothing but 1D Green's function of free 
particle in the half-space $x\ge 0$, contains all universal terms
of the small-$x,x'$ expansion (\ref{2.6}).
The remainder does not contain the universal terms and behaves like
\begin{equation} \label{3.4}
\delta G_z^{1{\rm D}}(x,x') = \left[ c_2^{1{\rm D}}(z) + 
{\rm i} \sqrt{z} \right] x x' + O[x x'(x^2+x'^2)] ,
\end{equation}
where the model-dependent $c_2^{1{\rm D}}(z)$ is defined by 
(\ref{2.5}) or (\ref{2.6}).
Substituting (\ref{3.3}) and (\ref{3.4}) in the basic formula (\ref{3.2}),
one gets
\begin{subequations} \label{3.5}
\begin{equation} \label{3.5a}
G_z^{3{\rm D}}({\bf r},{\bf r}') = - \frac{1}{4\pi} \left[
\frac{\exp\left( {\rm i}\sqrt{z}\vert {\bf r}-{\bf r}' \vert \right)}{
\vert {\bf r}-{\bf r}' \vert} - \frac{\exp\left( {\rm i}\sqrt{z}
\sqrt{(x+x')^2 + \rho_{\perp}^2} \right)}{\sqrt{(x+x')^2 + \rho_{\perp}^2}} 
\right] + \delta G_z^{3{\rm D}}({\bf r},{\bf r}') ,   
\end{equation}
where
\begin{equation} \label{3.5b}
\delta G_z^{3{\rm D}}({\bf r},{\bf r}')  =  - \frac{z}{2\pi}
\int_1^{\infty} {\rm d}s~ s \left[ c_2^{1{\rm D}}(z s^2)
+ {\rm i} \sqrt{z} s \right] I_0\left(\sqrt{z}
\rho_{\perp} \sqrt{s^2-1}\right)~ x x' + O \left[ x x'(x^2+x'^2) \right] .
\end{equation}
\end{subequations}
It can be easily shown that 
$c_2^{1{\rm D}}(z)+{\rm i}\sqrt{z} \sim O(1/z)$ as $\vert z\vert \to \infty$, 
so the integral in (\ref{3.5b}) has the necessary convergence property.
Consequently, the total $G_z^{3{\rm D}}(x,x';\rho_{\perp}\ne 0) \propto x x'$
does not exhibit universal expansion terms near the boundary, as is
intuitively anticipated.

When $\rho_{\perp}=0$, i.e. the two points ${\bf r}$ and ${\bf r}'$ 
lie on the same line perpendicular to the surface of the wall, relation
(\ref{3.2}) can be transformed to
\begin{equation} \label{3.6}
G_z^{3{\rm D}}(x,x';\rho_{\perp}=0) = G_0^{3{\rm D}}(x,x';\rho_{\perp}=0)
+ \frac{1}{4\pi} \int_0^z {\rm d}s~ G_s^{1{\rm D}}(x,x') .
\end{equation}
It is straightforward to show that the small-$x,x'$ expansion 
of $G_0^{3{\rm D}}$ is determined by the free-particle limit plus 
a model-dependent term $\sim O(x x')$.
From (\ref{2.6}), $G_s^{1{\rm D}}(x,x') = -x_< +O(x x')$.
Substituting this in (\ref{3.6}), one observes that
\begin{equation} \label{3.7}
G_z^{3{\rm D}}(x,x';\rho_{\perp}=0) = 
- \frac{x_<}{2\pi \vert x-x'\vert (x+x')} - \frac{z}{4\pi} x_<
+ O(x x') .
\end{equation}
The first two leading terms on the rhs of (3.7) are universal, 
the applied potential contributes starting from the term of order $O(xx')$.

We add that the 3D one-point quantity $\nu_z$, introduced in (\ref{1.6}),
equals the diagonal element of 1D Green's function associated with 
the potential $u(x)$, and therefore satisfies the inverse relation 
of type (\ref{1.8}), 
\begin{equation} \label{3.8}
z - u(x) = - \frac{1}{4}\frac{1}{\nu_z^2(x)} + \frac{1}{4}
\left[ \frac{\nu'_z(x)}{\nu_z(x)} \right]^2
- \frac{1}{2} \frac{\nu''_z(x)}{\nu_z(x)} ,
\end{equation}
and exhibits the small-$x$ expansion (\ref{2.5}).

\renewcommand{\theequation}{4.\arabic{equation}}
\setcounter{equation}{0}

\section{Radial Green's function}
We now consider a 3D quantum system with radial symmetry, 
confined to the domain $\Omega = \{ \vert {\bf r}\vert \le R \}$ 
(radius $R$ may be either finite or infinite).
The external potential $u(r)$ depends only on the magnitude 
$r$ of ${\bf r}$.
The radial problem can be reduced to 1D by factoring out the
angular dependence of the Hamiltonian eigenfunctions in terms
of the spherical harmonics.
Let us introduce for each angular momentum quantum number 
$l = 0, 1, 2,\ldots$ the 1D Green function associated with 
the Hamiltonian 
\begin{equation} \label{4.1}
{\bf \hat H}^{(l)} = - \frac{{\rm d}^2}{{\rm d} r^2}
+ \frac{l(l+1)}{r^2} + u(r) ,
\end{equation}
${\bf G}_z^{(l)} = (z - {\bf \hat H}^{(l)})^{-1}$, 
with zero b.c. at the origin $r=0$ and at the boundary $r=R$.
The total 3D radial Green's function is then expressible as
\begin{equation} \label{4.2}
G_z^{3{\rm D}}({\bf r},{\bf r}') = \sum_{l=0}^{\infty}
\frac{(2l+1)}{4\pi r r'} P_l(\cos\omega) 
G_z^{(l)}(r,r') ,
\end{equation}
where $\omega$ is the angle between ${\bf r}$, ${\bf r}'$ 
and $P_l$ denotes the Legendre polynomial of degree $l$.$^{24}$ 
The radial analogue of the 1D inverse profile Eq. (\ref{1.8})
for the diagonal elements of ${\bf G}_z^{(l)}$, 
$n_z^{(l)}(r) = G_z^{(l)}(r,r)$, reads
\begin{equation} \label{4.3}
z - u(r) - \frac{l(l+1)}{r^2} = - \frac{1}{4} \frac{1}{n_z^{(l)}(r)^2}
+ \frac{1}{4} \left[ \frac{{n_z^{(l)}}'(r)}{n_z^{(l)}(r)} \right]^2
- \frac{1}{2} \frac{{n_z^{(l)}}''(r)}{n_z^{(l)}(r)} ,
\end{equation}
with the obvious b.c. $n_z^{(l)}(0)=n_z^{(l)}(R)=0$.
According to (\ref{1.9}), the off-diagonal elements of ${\bf G}_z^{(l)}$ 
are expressible in terms of the diagonal ones as follows
\begin{equation} \label{4.4}
G_z^{(l)}(r,r') = \left[ n_z^{(l)}(r) \right]^{1/2} 
\left[ n_z^{(l)}(r') \right]^{1/2}
\exp \left[ {\rm sign} (r-r') \int_{r'}^r 
\frac{{\rm d}s}{2 n^{(l)}_z(s)} \right] .
\end{equation}

\subsection{Behavior close to the center}
Although the additional potential term $l(l+1)/r^2$ $(l\ne 0)$
is singular at $r=0$, the $r\to 0$ analysis of Eq. (\ref{4.3}) is similar 
to that of the 1D inverse profile relation (\ref{1.8}) close to a boundary 
(see Sec. 2).
Let us assume that the potential $u(r)$ is regular at the origin.
Inserting into (\ref{4.3}) the Taylor expansion of $n^{(l)}_z$ around $r=0$,
\begin{equation} \label{4.5}
n_z^{(l)}(r) = \sum_{n=1}^{\infty} c_n^{(l)} r^n , \quad \quad
c_n^{(l)} = \frac{1}{n!} \frac{{\rm d}^n n_z^{(l)}(r)}{
{\rm d}r^n}\big\vert_{r=0} ,
\end{equation}
one gets a sequence of equations for the coefficients $\{ c_n^{(l)} \}$
which implies
\begin{equation} \label{4.6}
n_z^{(l)}(r) \sim - \frac{r}{2l+1} + \delta_{l,0}
c_2^{(0)}(z) r^2 - \frac{2 z}{(2l-1)(2l+1)(2l+3)} r^3
\quad \quad {\rm as}\ r\to 0 .
\end{equation}
The leading terms of the expansion of Green's function
$G_z^{(l)}(r,r')$ in $r$ and $r'$ can be evaluated by inserting 
(\ref{4.6}) into (\ref{4.4}).
The total 3D Green's function (4.2) can be obtained
by applying the generating formula for Legendre polynomials,
with the result
\begin{equation} \label{4.7}
G_z^{3{\rm D}}({\bf r},{\bf r}') \sim - 
\frac{1}{4\pi \vert {\bf r}-{\bf r}' \vert}
+ \frac{c_2^{(0)}(z)}{4\pi} + \frac{z}{8\pi}
\vert {\bf r}-{\bf r}' \vert
\quad \quad {\rm as\ both}\  r, r' \to 0 .
\end{equation}
Here, the potential $u(r)$ and the presence of the boundary at $r=R$
are reflected only in the model-dependent coefficient $c_2^{(0)}(z)$.

\subsection{Behavior close to the boundary}
We introduce a new variable, the distance from the wall
$x=R-r$, and redefine $u(r)\to u(x)$, $n_z^{(l)}(r)\to n_z^{(l)}(x)$,
$G_z^{(l)}(r,r')\to G_z^{(l)}(x,x')$.
The total 3D Green's function (\ref{4.2}) is expressible simply as
\begin{equation} \label{4.8}
G_z^{3{\rm D}}(x,x';\omega) = \frac{1}{4\pi (1-x/R)(1-x'/R)}
\frac{1}{R} \sum_{l=0}^{\infty} \frac{(2l+1)}{R} P_l(\cos\omega)
G_z^{(l)}(x,x') .
\end{equation}
The "angular momentum" potential $l(l+1)/(R-x)^2$
is an analytic function of $x$ close to the boundary.
Provided that $u(x)$ is also regular at $x=0$, it holds
$G_z^{(l)}(x,x') \sim - x_<$ as $x, x' \to 0$. 
Inserting this into (\ref{4.8}), the 3D Green's function becomes 
proportional to the non-converging series 
$\sum_{l=0}^{\infty} (2l+1) P_l(\cos\omega)$ 
and we face the same problem as in the previous case of the
3D rectilinear boundary (Sec. 3).

Since we know how to solve the divergence problem for the
rectilinear hard wall, our strategy is first to reproduce
the result (\ref{3.2}), valid for the 3D rectilinear wall,
as the $R\to \infty$ limit of the relation (\ref{4.8}),
and then to get the leading $1/R$ correction due to 
the curvature of the wall surface.
We have treated the angular momentum potential $l(l+1)/(R-x)^2$
as a perturbative series in $1/R$, then used the standard
Green's function perturbation theory for ${\bf G}^{(l)}$
and finally performed a convenient continualization of 
the sum on the rhs of (\ref{4.8}) setting $l=kR-1/2$ ($k$ fixed)
and 
$$
P_l(\cos\omega) = J_0(k\rho_{\perp}) -
\frac{k\rho_{\perp}(x+x')}{2R} J_1(k\rho_{\perp})
+ O\left(\frac{1}{R^2}\right) .
$$ 
Here, $\rho_{\perp}$, defined by 
$\rho_{\perp}^2 = \vert {\bf r}-{\bf r}' \vert^2 - (x-x')^2$,
is the analogue of the perpendicular distance
$\vert {\bf r}_{\perp}-{\bf r}'_{\perp}\vert$
in the case of the rectilinear wall.
The final result is
\begin{equation} \label{4.9}
G_z^{3{\rm D}}(x,x';\rho_{\perp}) = \frac{1}{2\pi}
\int_0^{\infty} {\rm d}k~ k G_{z-k^2}^{1{\rm D}}(x,x')
J_0(k\rho_{\perp}) + \frac{1}{R} {\cal G}_z(x,x';\rho_{\perp})
+ O\left( \frac{1}{R^2} \right) ,
\end{equation}
where
\begin{eqnarray} \label{4.10}
{\cal G}_z(x,x';\rho_{\perp}) & = & \frac{(x+x')}{2\pi}
\int_0^{\infty} {\rm d}k~ k G_{z-k^2}^{1{\rm D}}(x,x') J_0(k\rho_{\perp})
\nonumber \\
& & - \frac{\rho_{\perp}(x+x')}{4\pi} 
\int_0^{\infty} {\rm d}k~ k^2 G_{z-k^2}^{1{\rm D}}(x,x') J_1(k\rho_{\perp})
\\ & & + 
\frac{1}{\pi}\int_0^{\infty} {\rm d}k~ k^3 J_0(k\rho_{\perp})
\int_0^{\infty}{\rm d}y~ G_{z-k^2}^{1{\rm D}}(x,y) y
G_{z-k^2}^{1{\rm D}}(y,x') . \nonumber
\end{eqnarray}
${\bf G}^{1{\rm D}}$ is the 1D Green's function corresponding
to the given potential $u(x)$, with the zero b.c. at $x=0$ and
the regularity condition at $x\to \infty$, 

The first term in (\ref{4.9}) is nothing but the exact result
(\ref{3.2}) valid for the $R\to\infty$ rectilinear hard wall.
Its expansion for small $x,x'$ coordinates was discussed in Sec. 3.
The next term ${\cal G}_z(x,x';\rho_{\perp})/R$ represents
the leading correction due to the curvature of the wall surface.
In the particular case of the zero potential, when ${\bf G}^{1{\rm D}}$
is given by (\ref{3.3}) with $\delta{\bf G}^{1{\rm D}}=0$,
the small-$x,x'$ expansion of ${\cal G}_z$ reads:
\begin{equation} \label{4.11}
{\cal G}_z(x,x';\rho_{\perp}\ne 0) =
\frac{x x' z}{4\pi} K_0(-{\rm i}\sqrt{z}\rho_{\perp})
+ O[x x' (x+x')] ,
\end{equation}
where $K_0$ is a modified Bessel function;$^{24}$
\begin{equation} \label{4.12}
{\cal G}_z(x,x';\rho_{\perp}=0) = \frac{x x'}{4\pi (x+x')^2}
- \frac{x x' z}{4\pi} \ln \left[ -{\rm i}\sqrt{z}(x+x') \right]
+ O(x x') .
\end{equation}
Worked-out examples with nonzero potentials tell us that for
$\rho_{\perp}\ne 0$ the leading term $\propto x x'$ 
on the rhs of (\ref{4.11}) depends on the applied potential,
while there is evidence that for $\rho_{\perp}=0$ the first
two leading terms on the rhs of (\ref{4.12}) are universal
(i.e. independent of the applied potential) and the potential
enters into the next term of order $O(x x')$.
From (\ref{4.12}), the one-point quantity of interest $\nu_z$  has
the following short-distance expansion from the wall
\begin{equation} \label{4.13}
\nu_z(x) = - x - \frac{1}{R} x^2 \ln \left( \frac{x}{x_0} \right)
+ O(x^2) ,
\end{equation}
where $x_0 = {\rm i}/\sqrt{z}$ is the length parameter.
We see that the nonzero curvature of the sphere surface induces
a non-analyticity, namely divergence of the second and higher-order 
derivatives of $\nu_z$ with respect to $x$ at the sphere surface $x=0$.
In the following section, we will prove the universality of the
logarithmic term in the expansion (\ref{4.13}) and derive its general
form for an arbitrarily shaped domain boundary.

\renewcommand{\theequation}{5.\arabic{equation}}
\setcounter{equation}{0}

\section{d-dimensional Green's function}
Let us now consider a $d$-dimensional quantum system with
the general Hamiltonian (\ref{1.1}) and Green's function
satisfying Eq. (\ref{1.5}).
Representing the $\delta$-function by
$\delta({\bf r}-{\bf r}') = \int \exp\left[{\rm i}{\bf k}\cdot
({\bf r}-{\bf r}')\right] {\rm d}{\bf k}/(2\pi)^d$,
${\bf k} = (k_1,\ldots,k_d)$, we rewrite (\ref{1.5}) as follows
\begin{equation} \label{5.1}
G_z({\bf r},{\bf r}') = [ z-u({\bf r})+\Delta_{\bf r}]^{-1}
\int \frac{{\rm d}{\bf k}}{(2\pi)^d} \exp\left[ {\rm i}
{\bf k}\cdot({\bf r}-{\bf r}') \right] .
\end{equation}
It is easy to derive the commutation rule
\begin{equation} \label{5.2}
[ z-u({\bf r})+\nabla_{\bf r}^2]^{-1} \exp({\rm i}{\bf k}\cdot {\bf r})
= \exp({\rm i}{\bf k}\cdot {\bf r}) [ z-u({\bf r})+
(\nabla_{\bf r}+{\rm i}{\bf k})^2]^{-1} .
\end{equation}
Consequently,
\begin{eqnarray} \label{5.3}
G_z({\bf r},{\bf r}') & = & 
\int \frac{{\rm d}{\bf k}}{(2\pi)^d} \exp\left[ {\rm i}{\bf k}\cdot
({\bf r}-{\bf r}')\right] [ z-u({\bf r})+ 
(\nabla_{\bf r}+{\rm i}{\bf k})^2]^{-1}  \nonumber \\ & = &
\int \frac{{\rm d}{\bf k}}{(2\pi)^d} \exp\left[ {\rm i}{\bf k}\cdot
({\bf r}-{\bf r}')\right] 
\left[ 1 + \frac{\nabla_{\bf r}^2 + 
2{\rm i}{\bf k}\cdot \nabla_{\bf r}}{z-u({\bf r})-k^2} \right]^{-1}
\frac{1}{z-u({\bf r})-k^2} .
\end{eqnarray}
The standard expansion of the inverse operator in (\ref{5.3}) then
results in an infinite series
\begin{eqnarray} \label{5.4}
& & \frac{1}{z-u({\bf r})-k^2} - \frac{1}{z-u({\bf r})-k^2}
(\nabla^2 + 2 {\rm i} {\bf k}\cdot \nabla) \frac{1}{z-u({\bf r})-k^2}
+ \nonumber \\ & &
\frac{1}{z-u({\bf r})-k^2} (\nabla^2 + 2 {\rm i} {\bf k}\cdot \nabla)
\frac{1}{z-u({\bf r})-k^2} (\nabla^2 + 2 {\rm i} {\bf k}\cdot \nabla)
\frac{1}{z-u({\bf r})-k^2} - \cdots
\end{eqnarray}
where we have dropped the subscript ${\bf r}$ from $\nabla$.
Writing the series (\ref{5.4}) formally as the sum
$\sum_n \alpha_n({\bf r},{\bf k})/[z-u({\bf r})-k^2]^n$,
we have at once
\begin{equation} \label{5.5}
\sum_{n=1}^{\infty} \frac{\alpha_n({\bf r},{\bf k})}{[z-
u({\bf r})-k^2]^n} = \frac{1}{z-u({\bf r})-k^2} -
\frac{1}{z-u({\bf r})-k^2} (\nabla^2 + 2{\rm i}{\bf k}\cdot \nabla)
\sum_{n=1}^{\infty}\frac{\alpha_n({\bf r},{\bf k})}{[z-
u({\bf r})-k^2]^n} .
\end{equation}
This differential equation implies the following recursion
for the coefficients $\{ \alpha({\bf r},{\bf k}) \}$:
\begin{eqnarray} \label{5.6}
\alpha_n & = & \delta_{n,1} - [ \nabla^2 \alpha_{n-1}
+ 2 {\rm i} {\bf k}\cdot \nabla \alpha_{n-1} ] \nonumber \\
& & - (n-2) [ 2{\rm i} \alpha_{n-2} {\bf k}\cdot \nabla u
+ 2\nabla \alpha_{n-2} \cdot \nabla u + \alpha_{n-2} \nabla^2 u] \\
& & - (n-2)(n-3) \alpha_{n-3} \vert \nabla u\vert^2 . \nonumber
\end{eqnarray}
Green's function (\ref{5.3}) is expressible in terms of the
coefficients $\{ \alpha_n \}$ as follows
\begin{equation} \label{5.7}
G_z({\bf r},{\bf r}') = \int \frac{{\rm d}{\bf k}}{(2\pi)^d}
\exp\left[ {\rm i}{\bf k}\cdot ({\bf r}-{\bf r}')\right]
\sum_{n=1}^{\infty} \frac{\alpha_n({\bf r},{\bf k})}{[z-
u({\bf r}) - k^2]^n} .
\end{equation}
Introducing spherical coordinates in ${\bf k}$-space, the integrals
on the rhs of (\ref{5.7}) can be evaluated by using the residue theorem
in the complex $k$-plane.

It is straightforward to retrieve the 1D profile relation (\ref{1.8})
by using the above scheme.
The calculations are more tedious in 3D.
Generating the coefficients $\{ \alpha_n \}$ from the recursion
(\ref{5.6}), formula (\ref{5.7}) gives, in the direct format,
\begin{eqnarray} \label{5.8}
{\rm i}\nu_z({\bf r}) & = & \frac{1}{2} \frac{1}{\sqrt{z-u({\bf r})}}
- \frac{1}{16} \frac{\Delta u({\bf r})}{[z-u({\bf r})]^{5/2}}
\nonumber \\
& & - \frac{5}{64}  
\frac{\vert \nabla u({\bf r}) \vert^2}{[z-u({\bf r})]^{7/2}} + 
\frac{1}{64} \frac{\Delta^2 u({\bf r})}{[z-u({\bf r})]^{7/2}}
+ O([z-u({\bf r})]^{-9/2}) .
\end{eqnarray}
The inversion of (\ref{5.8}) starting from the local-homogeneity 
reference $z-u({\bf r}) = - 1/[4\nu_z^2({\bf r})]$ results in 
the inverse profile equation
\begin{equation} \label{5.9}
z - u({\bf r}) = - \frac{1}{4\nu_z^2({\bf r})}
\left[ 1 - \vert \nabla \nu_z \vert^2 + 2 \nu_z \Delta \nu_z \right]
- \sum_{n=1}^{\infty} \varphi_n({\bf r}) .
\end{equation}
Here, the terms which involves just $2n$ $\nu_z$-functions are
grouped into $\varphi_n$.
In particular,
\begin{subequations} \label{5.10}
\begin{eqnarray}
\varphi_1 & = & \frac{1}{6} \left[ (\Delta \nu_z)^2 - 
\sum_{i,j=1}^3 (\partial_i \partial_j \nu_z)^2 \right] ,
\label{5.10a} \\
\varphi_2 & = & \frac{1}{6} \sum_{i,j,k}
\big[ 3 (\partial_i\partial_j \nu_z)^2 (\partial_k \nu_z)^2
+ 2 (\partial_i\partial_k \nu_z) (\partial_j\partial_k \nu_z)
(\partial_i \nu_z) (\partial_j \nu_z) \nonumber \\
& & \quad \quad -4(\partial_k^2 \nu_z)(\partial_i\partial_j \nu_z)
(\partial_i \nu_z) (\partial_j \nu_z) -
(\partial_i^2 \nu_z)(\partial_j^2 \nu_z)(\partial_k \nu_z)^2
\big] \nonumber \\
& + & \frac{1}{3} \nu_z \sum_{i,j,k} \big[
4(\partial_k^2 \nu_z) (\partial_i^2\partial_j \nu_z)(\partial_j \nu_z) - 
2(\partial_i^2\partial_j \nu_z)(\partial_j\partial_k \nu_z)
(\partial_k \nu_z) \nonumber \\ & & \quad \quad
- 2 (\partial_i\partial_j\partial_k \nu_z)
(\partial_i\partial_j \nu_z) (\partial_k \nu_z) 
-2 (\partial_i\partial_j \nu_z)(\partial_j\partial_k \nu_z)
(\partial_i\partial_k \nu_z) \nonumber \\
& & \quad \quad + 
(\partial_i\partial_j \nu_z)^2 (\partial_k^2 \nu_z) +
(\partial_i^2 \nu_z)(\partial_j^2 \nu_z)(\partial_k^2 \nu_z) \big]
\nonumber \\
& + & \frac{1}{6} \nu_z^2 \sum_{i,j,k} \big[
2(\partial_i^2\partial_j^2 \nu_z)(\partial_k^2 \nu_z)
- 2 (\partial_i\partial_j \nu_z)(\partial_i\partial_j\partial_k^2
\nu_z) \nonumber \\
& & \quad \quad + 
(\partial_i^2 \partial_k \nu_z)(\partial_j^2 \partial_k \nu_z)
-(\partial_i\partial_j\partial_k \nu_z)^2 \big] , \label{5.10b}
\end{eqnarray}
\end{subequations}
etc.
Note that the convergence of the direct series (\ref{5.8})
is restricted to the high-energy part of the spectrum,
$z-u({\bf r})>>1$, and to slowly varying potentials $u({\bf r})$.
These are the attributes of extended states, and, indeed,
$\nu_z$ given by (\ref{5.8}) exhibits a branch cut along real $z$-axis.
On the other hand, the inverse profile relation (\ref{5.9}),
being supplemented by the appropriate b.c. for $\nu_z$, holds
in the whole complex $z$-plane, including simple poles induced
by discrete localized states of the Hamiltonian spectrum.
This is an important feature of the inverse format.

As a check, we know from Sec. 3 that (\ref{5.9}) must reduce to the 
1D profile relation (\ref{3.8}) when both $u({\bf r})$ and $\nu_z({\bf r})$ 
are stratified in one dimension.
Under these circumstances $\varphi_1 = \varphi_2 = \cdots = 0$, 
and indeed (\ref{5.9}) becomes identical to (\ref{3.8}).

Like in the 1D case, the 3D inverse relation (\ref{5.9})
is very appropriate to describe the behavior of $\nu_z({\bf r})$
close to the boundary $\partial \Omega$, where $\nu_z\to 0$.
Provided that $u({\bf r})$ is regular at ${\bf r}\in \partial \Omega$,
the lhs of (\ref{5.9}) is regular at ${\bf r}\in \partial \Omega$
and so must be also the corresponding rhs.
The cancellation of the leading diverging terms on the rhs of 
Eq. (\ref{5.9}) is determined exclusively by the terms
$\propto 1/\nu_z^2$, i.e. by
\begin{equation} \label{5.11}
z - u({\bf r}) \sim - \frac{1}{4\nu_z^2({\bf r})}
\left[ 1 - \vert \nabla \nu_z \vert^2 + 2 \nu_z \Delta \nu_z \right] .
\end{equation}
This relation implies all universal terms of the $\nu_z$-expansion
around the boundary.

Let us analyze relation (\ref{5.11}) for the general potential $u({\bf r})$ 
and the general 3D domain $\Omega$ with a smooth 2D boundary 
$\partial \Omega$ of points ${\bf r}_0$ defined implicitly as follows
\begin{equation} \label{5.12}
\partial \Omega: \quad \quad \phi({\bf r}_0) = C_0 .
\end{equation}
The function $\phi$ is such that $C_0$ has the dimension of
length [for example, $\phi({\bf r}_0) = \sqrt{x_0^2+y_0^2+z_0^2}
(=R)$ for the sphere].
The set of points $\{ {\bf r}_0 \} \in \partial \Omega$ can be
parametrized by two curvilinear coordinates $\theta$ and $\varphi$, 
${\bf r}_0 = {\bf r}_0(\theta,\varphi)$, which are chosen to form
an orthogonal coordinate system.
To every point ${\bf r}\in \Omega$, we attach the number
$C=\phi({\bf r})$ (with the dimension of length) and introduce
the coordinate $\xi = C_0-C$.
The point ${\bf r}_0 \in \partial\Omega$ adjoint to a given ${\bf r}$
results as the intersection of the surface $\partial\Omega$ with
a curve which passes through ${\bf r}$ and simultaneously is perpendicular
at every point ${\bf r}'$ to the surface $\phi({\bf r}')=C'$.
The relationship between ${\bf r}$ and ${\bf r}_0$ reads
\begin{equation} \label{5.13}
{\bf r} = {\bf r}_0 + \sum_{n=1}^{\infty} {\bf u}_n({\bf r}_0) \xi^n ,
\end{equation}
where
\begin{subequations} \label{5.14}
\begin{eqnarray}
{\bf u}_1 & = & - 
\frac{\nabla \phi({\bf r}_0)}{\vert \nabla \phi({\bf r}_0)\vert^2} ,
\label{5.14a} \\
({\bf u}_2)_i & = & \frac{1}{2\vert \nabla \phi({\bf r}_0)\vert^4} 
\sum_j \partial_i\partial_j \phi({\bf r}_0) \partial_j \phi({\bf r}_0)
\nonumber \\
& & - \frac{\partial_i \phi({\bf r}_0)}{\vert \nabla \phi({\bf r}_0)\vert^6}
\sum_{j,k} \partial_j\partial_k \phi({\bf r}_0) \partial_j \phi({\bf r}_0) 
\partial_k \phi({\bf r}_0) , \label{5.14b}
\end{eqnarray}
\end{subequations}
etc.
The relationship (\ref{5.13}) determines ${\bf r}$ as a function 
of the curvilinear coordinates $\xi$, $\theta$ and $\varphi$, which are
orthogonal by construction.
Thus$^{24}$
\begin{subequations} \label{5.15}
\begin{eqnarray}
\vert \nabla \nu_z \vert^2 & = &
\frac{1}{h_{\xi}^2}\left( \frac{\partial \nu_z}{\partial \xi} \right)^2
+ \frac{1}{h_{\theta}^2}\left( \frac{\partial \nu_z}{\partial \theta} 
\right)^2
+ \frac{1}{h_{\varphi}^2}\left( \frac{\partial \nu_z}{\partial \varphi} 
\right)^2 , \label{5.15a} \\
\Delta \nu_z & = & 
\frac{1}{h_{\xi}^2}\left( \frac{\partial^2 \nu_z}{\partial \xi^2} \right)
+ \frac{1}{h_{\xi} h_{\theta} h_{\varphi}} 
\left( \frac{\partial}{\partial \xi} 
\frac{h_{\theta} h_{\varphi}}{h_{\xi}} \right) 
\frac{\partial \nu_z}{\partial \xi} + {\bf \hat \Phi}_{\theta,\varphi}
\nu_z , \label{5.15b}
\end{eqnarray}
\end{subequations}
where $h_{\xi}$, $h_{\theta}$ and $h_{\varphi}$ are metrical 
coefficients, and ${\bf \hat \Phi}_{\theta,\varphi}$ is the operator
containing derivatives only with respect to $\theta$ and $\varphi$.
It is useful to express the coefficients in (\ref{5.15}) in terms
of Cartesian coordinates:
\begin{eqnarray} \label{5.16}
\frac{1}{h_{\xi}^2} & = & \vert \nabla \phi({\bf r}) \vert^2
\nonumber \\ & = &
\vert \nabla \phi({\bf r}_0) \vert^2
- \frac{2\xi}{\vert \nabla \phi({\bf r}_0) \vert^2}
\sum_{i,j} \partial_i \partial_j \phi({\bf r}_0)
\partial_i \phi({\bf r}_0) \partial_j \phi({\bf r}_0) + O(\xi^2) ,
\end{eqnarray}
and
\begin{eqnarray} \label{5.17}
\frac{1}{h_{\xi} h_{\theta} h_{\varphi}} 
\left( \frac{\partial}{\partial \xi} 
\frac{h_{\theta} h_{\varphi}}{h_{\xi}} \right) & = &
-\Delta \phi({\bf r}) \nonumber \\
& = & - \Delta \phi({\bf r}_0) + O(\xi) .
\end{eqnarray}
Inspired by our previous result (\ref{4.13}) valid for the sphere,
the function $\nu_z$ is sought in the limit $\xi\to 0$ in the form
\begin{eqnarray} \label{5.18}
\nu_z({\bf r}) & = & c_1(\theta,\varphi) \xi +
c_2(\theta,\varphi) \xi^2 + \cdots \nonumber \\
& & + d_1(\theta,\varphi) \xi \ln(\xi/\xi_0)
+ d_2(\theta,\varphi) \xi^2 \ln(\xi/\xi_0) + \cdots ,
\end{eqnarray}
where the formal parameter $\xi_0 = {\rm i}/\sqrt{z}$ makes
the argument of the logarithm dimensionless.
The removal of diverging singularities on the rhs of Eq. (\ref{5.11}),
namely the vanishing of the coefficients attached to terms
$\ln^2\xi$, $\ln\xi$, $\xi^0$, $\xi\ln\xi$, $\xi\ln^2\xi$ and $\xi^1$
in the square bracket, fixes
\begin{equation} \label{5.19}
c_1 = - \frac{1}{\vert \nabla \phi({\bf r}_0)\vert}, \quad
d_1 = 0, \quad d_2 = - \frac{\Delta \phi({\bf r}_0)}{2 \vert \nabla 
\phi({\bf r}_0) \vert^3} .
\end{equation}
Consequently, one arrives at
\begin{equation} \label{5.20}
\nu_z({\bf r}) = - \frac{1}{\vert \nabla \phi({\bf r}_0)\vert} \xi -
\frac{\Delta \phi({\bf r}_0)}{2 \vert \nabla \phi({\bf r}_0) \vert^3} 
\xi^2 \ln(\xi/\xi_0) + O(\xi^2) .
\end{equation}

We present two examples.
For the sphere, $\phi({\bf r}_0) = \sqrt{x_0^2+y_0^2+z_0^2}
(=R)$, one has
\begin{equation} \label{5.21}
\nabla \phi({\bf r}_0) = \frac{{\bf r}_0}{\vert {\bf r}_0 \vert},
\quad \quad \Delta \phi({\bf r}) = \frac{2}{\vert {\bf r}_0\vert} .
\end{equation}
Since $\vert {\bf r}_0 \vert = R$ and $\xi = R - r$, one recovers
(\ref{4.13}) with $x\equiv \xi$.
For the cylinder infinite in the $z$-direction,
$\phi({\bf r}_0) = \sqrt{x_0^2+y_0^2} (=R)$, one has
\begin{equation} \label{5.22}
\nabla \phi({\bf r}_0) = \frac{{\bf r}_{\perp}}{\vert {\bf r}_{\perp} \vert},
\quad \quad \Delta \phi({\bf r}) = \frac{1}{\vert {\bf r}_{\perp}\vert} ,
\end{equation}
where ${\bf r}_{\perp}=(x_0,y_0,0)$.
Thus,
\begin{equation} \label{5.23}
\nu_z({\bf r}) = - \xi - \frac{1}{2 R} \xi^2 \ln(\xi/\xi_0)
+ O(\xi^2) ,
\end{equation}
with $\xi = R - \sqrt{x^2+y^2}$.

We would like to add that although $\xi$ has the dimension of length,
it does not represent in general the metric distance
$\tau = \vert {\bf r}-{\bf r}_0 \vert$.
Since one has
\begin{equation} \label{5.24}
\tau =  \frac{1}{\vert \nabla
\phi({\bf r}_0) \vert} \xi + O(\xi^2) ,
\end{equation}
Eq. (\ref{5.20}) can be expressed in terms of $\tau$ as follows
\begin{equation} \label{5.25}
\nu_z({\bf r}) = - \tau -
\frac{\Delta \phi({\bf r}_0)}{2\vert \nabla \phi({\bf r}_0) \vert}
\tau^2 \ln(\tau/\tau_0) + O(\tau^2)
\end{equation}
with the obvious definition of $\tau_0$.

As was mentionned in the Introduction, $\nu_z$, used through
relation (\ref{1.7}), is the crucial one-point quantity in
the density-functional formalism developed in Refs. 20-22.
From a general point of view it is desirable to have at one's
disposal also off-diagonal elements of Green's function.
The knowledge of $\nu_z$ is not sufficient to get the complete Green 
function within the present method 
(except of the case of the rectilinear hard wall).
However, the short-distance expansion of $\nu_z$ (\ref{5.25}),
derived with the aid of curvilinear coordinates which mimic
the global shape of the domain, depends only on the local
surface characteristic.
Since this one-point quantity is generated from Green's function
itself, it is reasonable to expect the local shape-dependence of the
latter as well.
This assumption, together with the explicitly worked-out examples
of the sphere (Sec. 4) and of the infinite cylinder, indicate
the following boundary behavior of Green's function.
For the considered pair of points ${\bf r}, {\bf r}' \in \Omega$
with the respective adjoint points ${\bf r}_0, {\bf r}'_0
\in \partial \Omega$, the perpendicular distance is defined by
$\rho_{\perp} = \vert {\bf r}_0 - {\bf r}'_0 \vert$ and
$\tau = \vert {\bf r}-{\bf r}_0 \vert$, 
$\tau' = \vert {\bf r}'-{\bf r}'_0 \vert$.
If $\rho_{\perp}\ne 0$, the 3D Green function is supposed to
exhibit the leading term $\propto \tau \tau'$ 
with a potential-dependent prefactor.
If $\rho_{\perp}=0$, the analogue of the sphere result (\ref{4.9})
reads
\begin{eqnarray} \label{5.26}
G_z^{3{\rm D}}(\tau,\tau';\rho_{\perp}=0) & = &
G_z^{3{\rm D}}(\tau,\tau';\rho_{\perp}=0)\big\vert_{\rm half-space}
+ \frac{\Delta \phi({\bf r}_0)}{2\vert \nabla \phi({\bf r}_0) \vert}
\Big\{ \frac{\tau \tau'}{4\pi (\tau+\tau')} \nonumber \\
& & - \frac{\tau \tau' z}{4\pi} \ln \left[ - {\rm i} \sqrt{z}
(\tau+\tau')\right] + O(\tau \tau') \Big\} + \cdots
\end{eqnarray}
The first term on the rhs of (\ref{5.26}) corresponds to Green's
function evaluated in presence of the rectilinear hard wall tangent
to the domain surface at point ${\bf r}_0$. 
The equation of the tangent plane is 
$\nabla \phi({\bf r}_0)\cdot ({\bf r}-{\bf r}_0)=0$.
The next term is the first correction due to the curvature of the wall
surface; the applied potential is supposed to contribute to the term
of order $O(\tau\tau')$.
A rigorous proof of the conjecture (\ref{5.26}) is a challenge
for the future.

In conclusion, we have shown that, close to a hard wall, time-independent
Green's functions exhibit a short-distance expansion which is to a
relatively high order universal, i.e. does not depend on the applied
potential, only the local shape of the wall is relevant.
The universal terms are analytic in 1D, and some of them are singular
in 3D provided that the boundary curvature is nonzero.
This information was obtained due to the inverse formalism developed
in this paper, and cannot be deduced directly from the definition
of Green's function.

\section*{Acknowledgments}
This work was supported by a grant from the National
Science Foundation and by Grant VEGA 2/7174/2.

\newpage

\noindent $^1$H. Weyl,
Nachr. Akad. Wiss. G\"ottingen, 110 (1911).

\noindent $^2$F. H. Brownell,
J. Math. Mech. {\bf 6}, 119 (1957).

\noindent $^3$M. Kac,
Amer. Math. Monthly {\bf 73}, 1 (1966).

\noindent $^4$T. Sunada,
Ann. Math. {\bf 121}, 169 (1985).

\noindent $^5$C. Gordon, D. L. Webb, and S. Wolpert,
Bull. Am. Math. Soc. {\bf 27}, 134 (1992).

\noindent $^6$T. A. Driscoll,
SIAM Rev. {\bf 39}, 1 (1997).

\noindent $^7$R. Balian and C. Bloch,
Ann. Phys. {\bf 60}, 401 (1970).

\noindent $^8$R. Balian and C. Bloch,
Ann. Phys. {\bf 64}, 271 (1971).

\noindent $^9$A. A. Actor,
Fortschr. Phys. {\bf 43}, 141 (1995).

\noindent $^{10}$R. Balian and C. Bloch,
Ann. Phys. {\bf 85}, 514 (1974).

\noindent $^{11}$E. Delabaere, H. Dillinger, and F. Pham,
J. Math. Phys. {\bf 38}, 6126 (1997).

\noindent $^{12}$R. W. Robinett,
J. Math. Phys. {\bf 39}, 278 (1998).

\noindent $^{13}$C. Zicovichwilson, J. H. Planelles,
and W. Jaskolski,
Int.J.Quan.Chem. {\bf 50}, 429 (1994).

\noindent $^{14}$W. Jaskolski,
Phys. Rep. {\bf 271}, 1 (1996).

\noindent $^{15}$M. E. Changa, A. V. Scherbinin, and V. I. Pupyshev,
J. Phys. B {\bf 33}, 421 (2000).

\noindent $^{16}$D. Bielinska-Waz, G. H. F. Diercksen,
and M. Klobukowski,
Chem. Phys. Lett. {\bf 349}, 215 (2001).

\noindent $^{17}$S. H. Patil,
J. Phys. B {\bf 35}, 255 (2002). 

\noindent $^{18}$P. Hohenberg and W. Kohn,
Phys. Rev. B {\bf 136}, 864 (1964).

\noindent $^{19}$W. Kohn and L. J. Sham,
Phys. Rev. A {\bf 140}, 1133 (1965).

\noindent $^{20}$J. K. Percus,
Int. J. Quan. Chem. {\bf 69}, 573 (1998).

\noindent $^{21}$L. {\v S}amaj and J. K. Percus,
J. Chem. Phys. {\bf 111}, 1809 (1999).

\noindent $^{22}$L. {\v S}amaj,
Int. J. Mod. Phys. B {\bf 15}, 2935 (2001).

\noindent $^{23}$E. N. Economou,
{\it Green's Functions in Quantum Physics}, 2nd ed.
(Springer-Verlag, Berlin, 1983).

\noindent $^{24}$I. S. Gradshteyn and I. M. Ryzhik,
{\it Tables of Integrals, Series, and Products}, 5th edn.
(Academic Press, London, 1994).

\end{document}